\documentclass[11pt]{amsart}

\usepackage{amsmath,amsthm,amssymb}
\usepackage{esint}             
\usepackage{bm}                
\usepackage{venndiagram}

\usepackage{a4wide}
\usepackage{fancyhdr}
\usepackage[breaklinks]{hyperref}
\usepackage{url}
\usepackage[numbers,square,sort&compress]{natbib}

\graphicspath{{./}{figs/}}
\usepackage{subcaption}
\usepackage[all]{xypic}
\usepackage{tikz}
\usetikzlibrary{arrows}
\usetikzlibrary{decorations.markings}
\usetikzlibrary{decorations.pathreplacing}
\usetikzlibrary{patterns}
\usetikzlibrary{shapes.geometric}
\usepackage{tikz-3dplot}

\usepackage{mhchem}
\usetikzlibrary{trees}

\makeatletter
\pretocmd\start@align{%
  \if@minipage\kern-\topskip\kern-\abovedisplayskip\fi
}{}{}
\makeatother

\usepackage{alltt}             
\usepackage{enumerate}
\usepackage{comment}

\usepackage{color}
\definecolor{darkblue}{rgb}{0, 0, .6}
\definecolor{grey}{rgb}{.7, .7, .7}
\hypersetup{colorlinks=true,linkcolor=darkblue,anchorcolor=darkblue,
  citecolor=darkblue,pagecolor=darkblue,urlcolor=darkblue,
  pdftitle={},pdfauthor={}
}

\theoremstyle{definition}

\def\<{\langle}
\def\>{\rangle}

\def\And{\wedge}
\def\Or{\vee}
\def\Not{\neg}

\def\longto{\longrightarrow}

\def\C{\mathbb{C}}
\def\F{\mathbb{F}}

\def\Q{\mathbb{Q}}
\def\R{\mathbb{R}}

\colorlet{lightred}{red!30!white}
\colorlet{lightblue}{blue!30!white}
\colorlet{lightyellow}{yellow!30!white}
\colorlet{keylime}{green!10!white}
\colorlet{darkgreen}{green!50!black}

\title{The case for algebraic biology: from research to education}

\author[M.~Macauley]{Matthew Macauley} \address{School of Mathematical and Statistical Sciences, Clemson University, Clemson, SC 29634} \email{macaule@clemson.edu}

\author[N.~Youngs]{Nora Youngs} \address{Department of Mathematics and Statistics, Colby College, Waterville, ME 04901} \email{nora.youngs@colby.edu}

\thanks{The first author was partially supported by Simons Foundation Grant \#358242. The second author is supported by the Clare Boothe Luce program.}



\begin{document}

\maketitle

\begin{abstract}
    Though it goes without saying that linear algebra is fundamental to mathematical biology, polynomial algebra is less visible. In this article, we will give a brief tour of four diverse biological problems where multivariate polynomials play a central role -- a subfield that is sometimes called \emph{algebraic biology}. Namely, these topics include biochemical reaction networks, Boolean models of gene regulatory networks, algebraic statistics and genomics, and place fields in neuroscience. After that, we will summarize the history of discrete and algebraic structures in mathematical biology, from their early appearances in the late 1960s to the current day. Finally, we will discuss the role of algebraic biology in the modern classroom and curriculum, including resources in the literature and relevant software. Our goal is to make this article widely accessible, reaching the mathematical biologist who knows no algebra, the algebraist who knows no biology, and especially the interested student who is curious about the synergy between these two seemingly unrelated fields.
\end{abstract}

\section{Introduction}

Nobody would dispute the fundamental role that linear algebra plays in applied fields such as mathematical biology. Systems of linear equations arise both as models of natural phenomena and as approximations of nonlinear models. As such, it is not hard to surmise that systems of nonlinear polynomials can also arise from  biological problems. Despite this, is still may come as a surprise to mathematicians and biologists alike when they first hear the term ``\emph{Algebraic Biology}.'' The reaction may be that of inquisitive curiosity, skepticism, or cynicism, as mathematicians have earned a reputation for occasionally making questionable abstractions and constructing frameworks that are perhaps too detached from reality to be more than just amusement. Some people who work in this field avoid the term ``algebraic biology'' for precisely this reason. Ironically, it actually might seem more reasonable to biologists, since most people without a degree in mathematics consider ``algebra'' to be a topic learned in middle or high school, and not a 400-level course involving abstract structures such as groups, rings, and fields. With this viewpoint, it would be only natural to ask ``\emph{If calculus can be an effective tool for tackling biological problems, why not algebra?}'' Regardless of a mathematician's reaction to the concept of algebraic biology, the aforementioned element of surprise can be attributed to the simple fact that abstract algebra is almost never taught in biology or biomathematics courses, nor are applications to biology usually presented in algebra courses. This is partially a statement about aspects of contemporary mathematical biology in the undergraduate and graduate curricula, and partially due to the actual nature of this multifaceted field. 

Though curricular reforms and improvements involving mathematical biology are certainly possible, we are not going to make the case in this article that algebra necessarily belongs as a fundamental pillar in a mathematical biology course. More realistically, problems from algebraic biology can make for fun, enriching, and interesting examples that can supplement such a course, and excite students who might like the topic but want to pursue something with direct applications. A good analogy of this is the RSA cryptosystem \cite{rivest1978method}. An entire number theory course can be taught without giving any applications. However, RSA is a fantastic self-contained topic that is not only captivating on its own, but gives a glimpse into how a traditional pure mathematical field such as number theory can be applied to real world problems. Being exposed to such an application will only broaden the appeal of the the mathematics behind the scenes, and it can motivate students to study it who might otherwise not give it a second thought. 

This is the role that we think algebraic biology should play in an undergraduate curriculum, though many may disagree. There are fundamental differences between the scope of algebraic techniques in biology and the ubiquity of those involving e.g., statistics or differential equations, which are more widespread, and for good reason. However, just as RSA can add spice to an ordinary number theory class, algebraic techniques for analyzing problems in the biological sciences can enhance both a mathematical biology class and an algebra class. In the process, it can give students a wider view of topics that they may want to pursue in their graduate studies.  

In this article, we will explore this further and dispel some myths. The first myth, which we have already alluded to, is that algebraic techniques are too detached from applications to be of any use. The second myth is that one needs a strong algebra background to study or teach these topics. While this may be true at the research level, it is absolutely false for both general undergraduate math majors and non-algebraically minded faculty alike. 

Our first task is to give a short summary of \emph{how} algebra actually arises in mathematical biology. Some readers may already know at least one of these topics somewhat well, but we will not assume this. It is also a hard truth in the biological world that ideas move quickly and the most modern methods all have a shelf life. Textbooks and papers quickly become antiquated as new technologies, theories, and computational power, become available. Not all, but certainly many decade-old books are as out-of-date to a biologist as a book on Visual BASIC or Pentium processors is to a computer engineer. One of the daunting goals in this paper is to give a snapshot of algebraic biology today in 2020, that will stand the test of time and remain relevant in 2030, and beyond. Of course, it is impossible to guarantee that this will actually happen, but it is nevertheless our objective. In the next section, we will present a few examples from algebraic biology at a high level -- only enough to convey the main ideas and framework to the reader, while making it clear how and why algebra is involved, with an assumption of only a minimal knowledge of algebra.

\section{Four algebraic problems arising from biological models}

The term ``algebra'' is quite broad and involves a plethora of structures, including groups, rings, fields, modules, vector spaces, and many more. Most of these have little to nothing to do with biology. The common algebraic theme in the problems that we will introduce are nonlinear \emph{multivariate polynomials}, which live in commutative rings. The branch of algebra that involves solving systems of such polynomials is called \emph{algebraic geometry}. In the problems we will encounter, computational techniques are particularly relevant. Polynomials arise in models of biological systems across a variety of frameworks, from classical differential equations, to Boolean networks, to statistical models in phylogenetics and genomics, to topics in neuroscience. The goal of this section is \emph{not} to provide a survey of these topics, or anything remotely close. However, we will briefly introduce them to give the reader the flavor of the biological questions and how (nonlinear) algebra arises. Our choice of these topics is motivated by several factors, including both the diversity of biological application, and their prominence in the mathematical biological community as of the writing of this article. Each of these topics should be thought of as a teaser. Like RSA over the past several decades, all of these topics have the potential to be the theme of a lecture or series of lectures on a creative application, aimed at undergraduates. This can be done in a course on algebra, mathematical biology, or even in a general-audience math club talk. We will provide and discuss introductory references for each topic for the reader who wants to learn more. 

\subsection{Biochemical reaction networks}

Consider the following simple biochemical reaction, where $A$, $B$, and $C$ are molecular species:
\[
\ce{$A+B$ <=>[$k_1$][$k_2$] $C$},\qquad \ce{$A$ ->[$k_3$] $2B$}.
\]
The constants $k_1$, $k_2$, and $k_3$ represent reaction rates. Let $x_1(t)$, $x_2(t)$, and $x_3(t)$ denote the concentrations of $A$, $B$, and $C$, respectively, where $t\in\R$. Without going into the details, the assumption of the fundamental laws of mass-action kinetics leads to the following systems of ordinary differential equations (ODEs)
\begin{align*}x_1'&=-k_1x_1x_2-k_3x_1+k_2x_3 \\ 
x_2'&=-k_1x_1x_2+k_2x_3+2k_3x_1 \\ 
x_3'&=k_1x_1x_2-k_2x_3.
\end{align*}
One of the most basic questions to ask when given a system of ODEs is: \emph{what are the steady-states?} Naturally, these can be found by setting each $x_i'=0$ and solving the resulting system of polynomial equations. From a biological perspective, we are really only interested in solutions that lie in the non-negative orthant of $\R^3$. However, wearing our mathematical hats, polynomials are usually easier to study over the complex numbers. In the language of algebraic geometry, for each fixed choice of parameters, the solutions to the system above form an \emph{algebraic variety} in $\C^3$. This can be found by defining the ideal  
\[
I=\big\<-k_1x_1x_2-k_3x_1+k_2x_3,\;-k_1x_1x_2+k_2x_3+2k_3x_1,\;k_1x_1x_2-k_2x_3\big\>
\]
and using a computer algebra package such as Macaulay2 \cite{M2} or Singular \cite{singular} to compute a Gr\"obner basis. One does not need to know what an ideal, Gr\"obner basis, or algebraic variety is to be able to carry out these steps. Though this may seem unsettling at first, scientists routinely treat solvers as black boxes when they solve an ODE numerically, use a linear model to analyze data, or solve a multiobjective optimization problem. Though the mathematics behind the scenes for most of these is beyond the undergraduate level, the utility and accessibility of the methods is usually not. That said, it is still nice to have some high-level idea of what the black box is doing, even if purely for consolation. A reader who is unfamiliar with the algebra can think of our particular black box as doing what Gaussian elimination does to solve systems of linear equations, but instead for systems of polynomials. The terms ideal, Gr\"obner basis, and algebraic variety are loosely analogous to the concepts of a vector space, a special vector space basis, and the solution space.

Of course, this is just the tip of the mathematical iceberg. There are plenty of research projects available for people more interested in theoretical aspects than the actual application to specific biological systems. The 2019 book \emph{Foundations of Chemical Reaction Network Theory} by Martin Feinberg, one of the pioneers of this field, is a great starting point for readers interested in learning more \cite{feinberg2019foundations}.

\subsection{Boolean models of molecular networks}

Our second example of algebra arising in a biological problem comes from modeling molecular networks with Boolean variables. As an alternative to quantifying the concentrations of the relevant gene products and modeling them with a system of nonlinear differential equations, it has become popular to qualitatively describe the variables as high vs. low (or present vs. absent), and describe their interactions with Boolean logic. Recall that $\And$, $\Or$, and $\neg$ represent logical AND, OR, and NOT, respectively. For example, suppose that for gene $A$ to be transcribed, enzyme $B$ must be present, and repressor protein $C$ must be absent. This can be expressed as $A(t+1)=B(t)\And\neg C(t)$. The following Boolean model of the lactose (\emph{lac}) operon in \emph{E. coli} is from \cite[Chapter 1]{robeva2013mathematical}:
\begin{align*}
x_1(t+1)&=\Not{G_e}\And(x_3(t)\Or L_e) \\
x_2(t+1)&=x_1 \\
x_3(t+1)&=\Not{G_e}\And[(L_e\And x_2(t))\Or(x_3(t)\And\Not{x_2(t)})].
\end{align*}
The \emph{lac} operon was the first, and remains one of the most well-studied gene networks in molecular biology, and the scientists who discovered it won a Nobel Prize for their work \cite{jacob1961genetic}. Here, time is assumed to be discretized, and $x_1(t)$, $x_2(t)$, and $x_3(t)$ are Boolean functions that represent the presence or absence of intracellular mRNA, translated proteins, and lactose, respectively. There are also two parameters, $L_e$ and $G_e$, representing extracellular lactose and glucose, respectively. These can be thought of as constants, because they change on a much larger time-scale than the three variables do. Like we did with our system of ODEs from a biochemical reaction network, we can ask about the steady states of this model. These can be found by setting $x_i(t+1)=x_i(t)$, and solving the resulting system. This is easiest by first converting the Boolean expressions into polynomials: $x\And y$ is $xy$, $x\Or y$ is $x+y+xy$, and $\Not{x}$ is $1+x$. This gives the following system of polynomials over $\F_2=\{0,1\}$:
\begin{align*}
(1+G_e)(x_3L_e+x_3+L_3)+x_1&=0 \\
x_1+x_2&=0 \\
(1+G_e)(x_2L_e+x_3(1+x_2))+x_3&=0.
\end{align*}
For each choice of parameters, a computer algebra package can find the solution by computing a Gr\"obner basis of the ideal
\[
I=\big\<(1+G_e)(x_3L_e+x_3+L_e)+x_1,\;
x_1+x_2,\;(1+G_e)(x_2L_e+x_3(1+x_2))+x_3\big\>.
\]
Naturally, this time everything needs to be done over the finite field $\F_2$. For more details on this at an undergraduate level, see the first several chapters of the 2013 book \emph{Mathematical Concepts and Methods in Modern Biology} by Robeva and Hodge \cite{robeva2013mathematical}. It should be noted that in some cases, the Boolean assumption is too restrictive, and modelers use more than just two states; sometimes these are called \emph{logical models} \cite{thomas2006biological}, \emph{local models} \cite[Chapter 4]{robeva2018algebraic}, or \emph{algebraic models} \cite{macauley2020algebraic}.

We chose the aforementioned examples -- one from an ODE model of a biochemical reaction network, and another from a toy model of the \emph{lac} operon -- because they are quite simple but still illustrate the main concepts. Published models from the literature are typically much larger and more complex. A Boolean model of the \emph{lac} operon in \cite{stigler2012regulatory} has 10 variables and 3 parameters, but the fundamental mathematical ideas are still the same. Most biochemical reaction networks in the literature also contain more than three species. For example, the authors in \cite{gross2016algebraic} study a model of the Wnt signaling pathway, which leads them to system of 19 ODEs with 31 rate constants, and 5 additional parameters from conservation laws. For a fixed choice of these parameters, the steady-state variety lives in $\C^{19}$. In the generic case, this variety lives in the algebraic closure of a rational function field, i.e., $K=\overline{\Q(k_1,\dots,k_{31},c_1,\dots,c_5)}$. They show that it has nine zeros (fixed points). In addition to shedding light on the biological questions, these problems generate many interesting mathematical problems. As Bernd Sturmfels has prominently said, \emph{biology can lead to new theorems} \cite{sturmfels2005can}.

\subsection{Algebraic statistics and genomics}

The field of algebraic statistics emerged around the turn of the 21st century \cite{diaconis1998algebraic}, and the term itself was coined in the year 2000 with the publication of a book bearing its name \cite{pistone2000algebraic}. Since then, it has blossomed into an active field of research. The basic idea is to use problems from algebraic geometry to study certain problems in statistics. Just as most biological problems are not well-suited for algebraic tools, most statistical problems are not either. Those that are typically involve discrete random variables, and the algebra comes into play when these depend on parameters in a polynomial fashion. The most common examples of this involve genomics and phylogenetics, where the nucleic acid bases (A, C, G, T) can be treated as discrete random variables. Biological questions that one might ask are how to identify regions in a genome with a higher concentration of CpG dinucleotides (often an indicator for coding regions), what is the most likely sequence of mutations between species between two samples, or what phylogenetic tree best represents the evolution of a collection of species. 

For a toy example that illustrates the connection to algebra, consider a simple evolutionary relationship of two species and their most common ancestor, and fix a particular base in the genome at a site that all three species share in a mutual alignment. The Jukes-Cantor model of evolution \cite{jukes1969evolution} assumes that the probability of a mutation at that site is $\alpha$, and therefore the probability the base not changing is $1-3\alpha$. We can express this with a tree, where the ancestor is the root, and the descendants are at the leaves. The Jukes-Cantor constants might be different for each species, so we will denote them by $\alpha$ and $\beta$, respectively. If we assume that A, C, G, and T are equally likely at this site in the ancestor genome, then we can compute the probability of all 16 cases for the leaves. The following example illustrates this. 

\begin{minipage}{0.25\linewidth}
\begin{center}
\begin{tikzpicture}[level distance=1.5cm,
  level 1/.style={sibling distance=2.3cm},
  level 2/.style={sibling distance=1.5cm}]
  \node {ancestor}
    child {node {human} edge from parent node[left,draw=none] {$\alpha$\,} 
    }
    child {node {chimp} edge from parent node[right,draw=none] {\;$\beta$}
    };
\end{tikzpicture}
\end{center}
\end{minipage}%
\hfill%
\begin{minipage}{0.7\linewidth}
\begin{align*} 
      P(AC)&=P\bigg(\hspace{-1mm}\begin{tikzpicture}[baseline={([yshift=-20pt]current bounding box.north)},shorten >= -2pt, shorten <= -2pt]
      \node (l) at (0,0) {\footnotesize $A$};
      \node (r) at (.9,0) {\footnotesize $C$};
      \node (rt) at (.45,.75) {\footnotesize $A$};
      \draw (rt) to (l); \draw (rt) to (r);
      \end{tikzpicture}\hspace{-1mm}\bigg)
    +P\bigg(\hspace{-1mm}\begin{tikzpicture}[baseline={([yshift=-20pt]current bounding box.north)},shorten >= -2pt, shorten <= -2pt]
      \node (l) at (0,0) {\footnotesize $A$};
      \node (r) at (.9,0) {\footnotesize $C$};
      \node (rt) at (.45,.75) {\footnotesize $G$};
      \draw (rt) to (l); \draw (rt) to (r);
      \end{tikzpicture}\hspace{-1mm}\bigg)
     +P\bigg(\hspace{-1mm}\begin{tikzpicture}[baseline={([yshift=-20pt]current bounding box.north)},shorten >= -2pt, shorten <= -2pt]
      \node (l) at (0,0) {\footnotesize $A$};
      \node (r) at (.9,0) {\footnotesize $C$};
      \node (rt) at (.45,.75) {\footnotesize $C$};
      \draw (rt) to (l); \draw (rt) to (r);
      \end{tikzpicture}\hspace{-1mm}\bigg)
      +P\bigg(\hspace{-1mm}\begin{tikzpicture}[baseline={([yshift=-20pt]current bounding box.north)},shorten >= -2pt, shorten <= -2pt]
      \node (l) at (0,0) {\footnotesize $A$};
      \node (r) at (.9,0) {\footnotesize $C$};
      \node (rt) at (.45,.75) {\footnotesize $T$};
      \draw (rt) to (l); \draw (rt) to (r);
      \end{tikzpicture}\hspace{-1mm}\bigg) \\
      &=\frac{1}{4}(1-3\alpha)\beta+\frac{1}{4}\alpha\beta+\frac{1}{4}\alpha(1-3\beta)+\frac{1}{4}\alpha\beta
      =\frac{1}{4}(\alpha+\beta-\alpha\beta).
      \end{align*}
\end{minipage}

\vspace{5mm}

      It is easy to verify that similarly, $P(AA)=\frac{1}{4}(1-3\alpha)(1-3\beta)+\frac{3}{4}\alpha\beta=3\alpha\beta+\frac{1}{4}(1-3\alpha-3\beta)$. The space of possible probabilities can be described by a mapping
      \[
      \varphi\colon\R^2\longto\R^{16},\qquad\varphi\colon(\alpha,\beta)\longmapsto\big(P(AA),P(AC),\dots,P(TT)
      \big).
      \]
      If we have an $n$-leaf tree, then things get complicated quickly. Besides the numbers rapidly increasing, we also have to consider the different tree topologies. If we fix an $n$-leaf binary tree $T$, which has $m=2n-2$ edges and hence $m$ probability parameters, we get a mapping $\varphi\colon\R^m\to\R^{4^n}$. The image of this map, intersected with the $d=4^n-1$ dimensional simplex $\Delta_d$ (because we require the coordinates to represent probabilities) is called the \emph{phylogenetic model}, $\mathcal{M}_T\subseteq\R^{4^n}$.
      
      The skeptical reader may wonder about the feasibility of this approach, given how quickly the size of $4^n$ grows. Fortunately, work has been done on analyzing the case of trees with $n=4$ leaves, and then patching them together to learn about the general case. Given the set of points $\mathcal{M}_T\subseteq\R^{4^n}$,
      a typical object for an algebraic geometer to look at is the set (ideal) of polynomials $f$ in $\R[x_1,\dots,x_{4^n}]$ such that $f(p)=0$ for all $p\in\mathcal{M}_T$. This is called the \emph{ideal of phylogenetic invariants}, $I_T$. Given any ideal, it is natural to look at the corresponding variety, which is the set of points $p\in\R^{4^n}$ such that $f(p)=0$, for all polynomials $f\in I_T$. This is called the \emph{phylogenetic  variety} of $T$, denoted $V_T$. 
      
 Hopefully the reader can now see how polynomials arise in this type of work. In fact, many of the advanced statistical models used by computational biologists, such as Markov models and graphical models, can be viewed as  algebraic varieties \cite{pachter2007mathematics}. To learn more about these topics the interested reader can consult books such as Allman/Rhodes (basic undergraduate introduction to phylogenetics) \cite{allman2004mathematical}, Pachter/Sturmfels \cite{pachter2005algebraic}, or Sullivant \cite{sullivant2018algebraic} (both graduate-level algebraic statistics).

\subsection{Place fields in neuroscience}

\begin{figure}[!ht]
\begin{center}
\includegraphics[width=.35\textwidth]{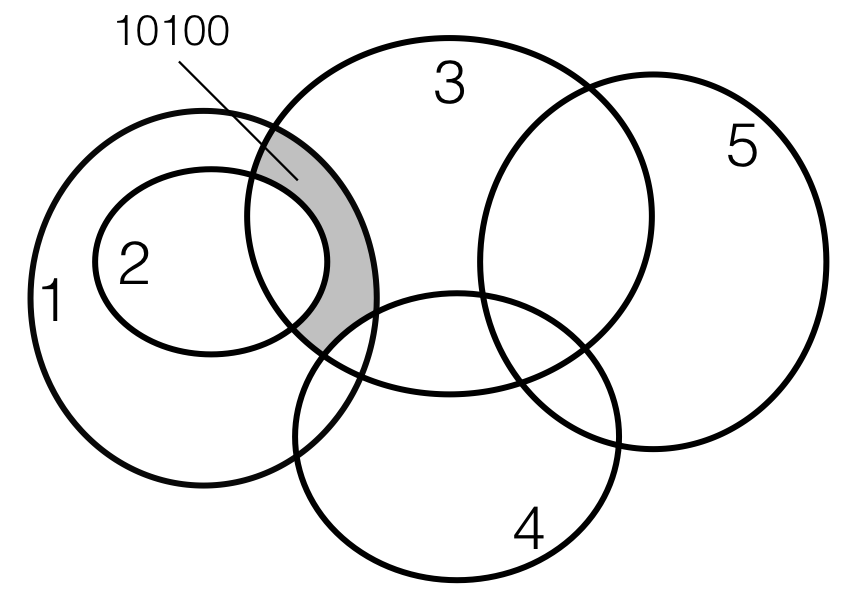}
\end{center}
\caption{An example of five place fields. The shaded region is described by the codeword $\mathbf {c}=10100$.}\label{fig:place-fields}
\end{figure}

Our last topic comes from a problem in the vast field of neuroscience. Individuals are able to perceive sensory input such as light, color, motion, and location via neurons in the brain that respond to specific stimuli. Experiments have shown that certain neurons called \emph{place cells} fire based on the location of an animal in its environment \cite{okeefe1971hippocampus,yartsev2013representation}. As an animal moves around to different locations, different subsets of neurons fire. The region that causes a specific neuron to fire is called its \emph{place field}, and a single environment will generally exhibit many overlapping place fields. A cartoon of such place fields in two dimensions is shown in Figure~\ref{fig:place-fields}.
Notice that each region in the ``Venn diagram'' of place fields determines a binary string indicating which neurons should fire therein, called its associated \emph{codeword}. The collection of all strings possible within the arrangement is the \emph{code} of the set of place fields. It is a simple matter to construct the code from the place fields. However, the \emph{inverse problem} of deducing an appropriate set of regions corresponding to the firing of neurons given experimental data is not only much more difficult, but more scientifically relevant, because the structure of the place fields are not known \emph{a priori}. Not all codes correspond to a natural collection of place fields. For example, after a few minutes of trying, it is not hard to convince oneself that the code $\mathcal{C}=\{000,100,010,101,110\}$ \emph{cannot} be realized by a collection of convex open place fields. This fails even in higher dimensions, as the place field for Neuron 3 would need to be split between that for Neuron 1 and Neuron 2 \cite{curto2013neural}.

A natural approach for this problem involves algebraic geometry. To motivate this, consider a codeword $\mathbf{c}=010$, which can be encoded by its characteristic polynomial $p_{\mathbf{c}}=(x_1-1)x_2(x_3-1)$; note that $p_{\mathbf{c}}(\mathbf{c})=1$, and $p_{\mathbf{c}}(\mathbf{c}')=0$ for all other length-3 codewords $\mathbf{c}'$. This polynomial is called a \emph{pseudomonomial} because by viewing $x_1-1=x_1+1$ over $\F_2$ as ``NOT $x_1$'' in the Boolean sense, it can be written as a ``monomial with bars allowed'', e.g., $p_{\mathbf{c}}=\overline{x_1}x_2\overline{x_3}$. One can define the so-called \emph{pseudomonomial ideal}
\[
I_{\mathcal{C}}:=\big\{f\in\F_2[x_1,\dots,x_n]\mid f(\mathbf{c})=0\;\text{for all }\mathbf{c}\in\mathcal{C}\big\},
\]
and the \emph{neural ring} is simply the quotient of $\F_2[x_1,\dots,x_n]$ by this ideal. Another object of interest is the \emph{neural ideal} $J_\mathcal{C}$, which is generated by the characteristic functions of the non-code words, i.e., words \emph{not} in $\mathcal{C}$. These ideals describe the neural code completely, and understanding the generators of such an ideal can give insight into both the possibility of building an arrangement of place fields which exhibits the code, and what is the proper dimension in which to do so \cite{curto2013neural, curto2019algebraic,garcia2018grobner}. Once again, this is an example of new mathematical problems and theorems inspired by biology. However, as this work originated from a biological problem, there are also many questions about how to deal with noisy data, and robustness of these techniques -- questions that may not have arisen if these problems had originated organically within the algebra community. Of the four topics summarized in this paper, this is the newest, and it was only  first published in 2013 \cite{curto2013neural}. However, it has exploded in popularity since, especially due to all of the unexplored mathematical problems that have arisen. Readers wishing to learn more about this should consult the survey article published in \cite[Chapter 7]{robeva2018algebraic}, and the references therein. Also, in a forthcoming paper, the first author and Raina Robeva will write a more thorough survey about pseudomonomials in algebraic biology, and how they appear in various different topics -- not only in the analysis of combinatorial neural codes, but also involving one of our other four topics, algebraic models of gene networks.

\section{A brief history of algebraic biology}

Methods and models in mathematical biology that involve networks, discrete mathematics, and algebraic techniques often fall under different overlapping umbrellas -- systems biology, bioinformatics, discrete mathematical biology, algebraic biology, algebraic systems biology, and so on. We tend to be rather loose with the term ``algebraic biology'' and what it encompasses because there is a large gray area. For example, one can build a Boolean model of a gene regulatory network, and analyze it without bringing algebra into the picture. In fact, the vast majority of Boolean network research does not involve algebra at all -- much of it comes from disciplines such as computer science \cite{crama2010boolean}, engineering \cite{cui2010complex}, or physics \cite{drossel2009random,wang2012boolean}. However, algebra still lies behind the scenes, whether or not it is utilized. We tend to err on the inclusive side of what fits under the big tent of algebraic biology.

To understand where algebraic biology  may be going and in what context it belongs in the classroom, it is necessary to look back and see how we got to where we are today. In the late 1960s and early 1970s, modeling biological systems with Boolean functions was pioneered independently by two theoretical biologists, Stuart Kauffman in North America, and Ren\'e Thomas in Europe. These ideas percolated through countless collaborators, especially their students and postdocs, through multiple academic generations, and they remain a popular research topic globally today. However, it was not until the turn of the century that these ideas took a more algebraic fork, much of which was inspired by Reinhard Laubenbacher and his large academic tree. An anecdote of this shift is described in the 2009 American Mathematical Monthly article \emph{Computer algebra and systems biology} by Laubenbacher and Sturmfels \cite{laubenbacher2009computer}. 

The theory of chemical reaction networks emerged around the same time as Boolean networks, as it was pioneered by Martin Feinberg, Friedrich Horn, and Roy Jackson in the early 1970s. The use of computational algebra was instigated by Karin Gatermann's 2001 book titled ``Computer algebra methods for equivariant dynamical systems'' \cite{gatermann2007computer}, and it remains an active area of research today, especially among the academic family trees and collaborators of Feinberg, Sturmfels, and Alicia Dickenstein, among others. Feinberg published an introductory book \emph{Foundations of chemical reaction network theory} in 2019 \cite{feinberg2019foundations}.

Algebraic techniques have sporadically been used in statistical problems since at least the 1940s \cite{wilks1946sample,votaw1948testing}, but the term ``\emph{Algebraic Statistics}'' was not coined until 2000, when it appeared in the title of the book \emph{Algebraic Statistics: Computational Commutative Algebra in Statistics} by Giovanni Pistone, Eva Riccomagno, and Henry Wynn \cite{pistone2000algebraic}. It is up for debate as to how much a catchy name catalyzed interest in this new exciting budding field, but it nevertheless has taken off since then. A graduate topics class at UC Berkeley led to the popular 2005 ``purple book'' \emph{Algebraic Statistics for Computational Biology} by Sturmfels and Lior Pachter, with a number of chapters contributed by graduate students and postdocs. New journals sprang up, such as the \emph{Journal of Algebraic Statistics} (2010), and the \emph{SIAM Journal on Applied Algebra and Geometry} (2017). In 2018, Seth Sullivant published the first broad introductory (graduate-level) book on algebraic statistics \cite{sullivant2018algebraic}.

Finally, neural rings first appeared in the literature in 2013, in a series of papers by Carina Curto, Vladimir Itskov, and their students, postdocs, and collaborators \cite{curto2013combinatorial,curto2013neural,youngs2014neural}. It has since taken off and spurred an active area of research, because aside from the original motivation of neuroscience, the neural rings and pseudomonomial ideals were new mathematical objects with a theory waiting to be developed, involving questions from combinatorial algebra, geometry, and topology. 

Though several of these four areas date back to the late 1960s or early 1970s, the strong connection to algebraic geometry did not really take off until the turn of the century. As such, some of the foundational books on these topics are only now starting to appear.  As with many new fields, compilations of loosely related book chapters precede broad introductory texts such as \cite{feinberg2019foundations, sullivant2018algebraic}. The purple book of Pachter and Strumfels is one such example in algebraic statistics. Raina Robeva has edited a series of three books on discrete and algebraic methods in mathematical biology, published in 2013, 2015, and 2018, respectively \cite{robeva2013mathematical,robeva2013mathematical,robeva2018algebraic}. They include 36 chapters that vary in difficulty from the introductory undergraduate level to more advanced graduate levels. Since the chapters are written by many different authors, there are at times overlaps and varying notations. However, all of these are good starting points for teaching some of these topics at the undergraduate or graduate levels, and a number of faculty members have used parts of these books for classes, especially the first one. Other good introductory chapters can be found in edited volumes on related topics, both in mathematical biology \cite{jonoska2013discrete} and in applications of subjects such as discrete mathematics \cite{harrington2017algebraic} and polynomial systems \cite{cox2020applications}.

As previously stated, our choice of four topics to highlight in this article was driven by the fact that they have remained active areas of research. This is by no means a comprehensive list of past or current topics in algebraic biology. A series of conferences titled \emph{Algebraic Biology} were held in 2005, 2007, 2008, and 2010, with the last one called \emph{Algebraic and Numeric Biology}. In the early 2000s, theoretical chemist and professor emeritus Michael Barnett published several articles on computer algebra in the life sciences, foreseeing the relevance to computational algebra and symbolic computation that was only starting to blossom. The large field of phylogenetics employs some algebraic tools and methods, though much of this can be considered part of algebraic statistics. Phylogenetics was one of the catalysts of the now thriving 21st century field of tropical geometry, a ``piecewise linear'' variant of algebraic geometry, with a number of diverse applications \cite{maclagan2015introduction}. The relatively new area of \emph{topological data analysis} has emerged over the past decade, and this involves using algebraic topology -- homology groups, in particular -- to analyze the shape of clouds of data \cite{wasserman2018topological}. A number of these applications are problems in the biological sciences, and algebra is fundamental to the topological methods \cite{rabadan2019topological}. As such, this field certainly belongs under the broad umbrella of algebraic biology. Heather Harrington at Oxford leads a large research group branded as \emph{algebraic systems biology}, and much of their work involves algebraic statistics, chemical reaction networks, and topological data analysis. 

Though the footprint of algebraic biology is large and established, not everyone is in agreement on what to call it, or whether it the broad topics should even be given a collective name. The authors of this article like the term \emph{Algebraic Biology} not only because it is catchy, but because it is both informative to mathematicians about the scope while not being off-putting to non-mathematicians, due the word ``algebra'' being fairly innocuous. Sometimes, depending on the scope, it may be desirable to use additional terms, and get names such as \emph{Computational Algebraic Biology}, or \emph{Algebraic Systems Biology}. Having a memorable title can help unify and strengthen an emerging field, like what was done with algebraic statistics and topological data analysis around the turn of the century, energizing researchers and prospective students alike.

\section{Algebraic biology in education}

In contrast to the many people who go to graduate school with plans to study mathematical biology, most people who work in algebraic biology seem to end up there by accident. Some came from pure math and stumbled upon applications; others ended up there because they found an advisor they liked who worked in that area. As these areas grow, this is perhaps gradually changing, but slowly. 

There are many reasons why most of the mathematical biology content that students see involves classical techniques such as ordinary differential equations (ODEs). For one, the field is older and more developed. It is also broader, as there are many natural examples to put in an ODE course, such as logistic growth, the SIR model, and nonlinear models such as competing species or predator-prey. Certain fundamentals are ubiquitous throughout the biological sciences -- the need to analyze data, relationships between rates of change, and the need to approximate relationships with linear functions. Thus, it should come as no surprise to see statistics, calculus, differential equations, and linear algebra dominate the mathematical biology community. Algebraic techniques, models, and methods comprise a niche in biology, much like they do in statistics. This makes it less likely to see algebraic biology taught as a full semester class, though it certainly can be and has been done. Entire topics classes have been taught on the topics we have discussed -- Boolean models, biochemical reaction networks, algebraic statistics, and phylogenetics. Often, such classes are at the graduate level, but many are still very accessible to undergraduates. Additionally, as previously stated, select topics on their own can make a great supplement to both a traditional mathematical biology, modeling, or even abstract algebra class. It is also a good source of open-ended problems that are accessible for undergraduate research. 

One challenge to teaching these materials in the classroom is the lack of standard textbooks, though this is changing with the recent publication of introductory texts on fields such as biochemical reaction networks \cite{feinberg2019foundations} and algebraic statistics \cite{sullivant2018algebraic}. The three aforementioned books on various discrete and algebraic methods in mathematical biology \cite{robeva2013mathematical,robeva2015algebraic,robeva2018algebraic} edited or co-edited by Robeva are all freely available online with a ScienceDirect subscription, which many institutions have. Three out of the four topics covered in this article are contained in at least one chapter, with the exception being algebraic statistics. This is arguably the most advanced topic of the four, because algebraic geometry plays a central role. The first author of this article regularly teaches a course on mathematical modeling and divides it roughly into thirds: continuous, discrete, and stochastic methods, with algebra arising in these last two sections. He and Robeva have co-organized three workshops about teaching discrete and algebraic methods in mathematical biology to undergraduates, and the last two of Robeva's books partially came out of these workshops. A number of participants in these workshops have experimented with teaching topics in their own classes. The second author of this paper has used the most recent of these texts \cite{ robeva2018algebraic} to generate a self-contained unit of biological applications in a first abstract algebra course.

Another challenge to introducing algebraic methods into the classroom is the lack of established software tools available for analysis. Corporate packages such as Matlab, Maple, and Mathematica do not have the need or incentive to implement relevant algorithms. The result is that most software tools arise out of research groups at academic institutions, which are fluid and sometimes dependent on grant funding. One of the most successful and well-known such software packages for Boolean models is the freely available Gene Interaction Network simulation (GINsim) software \cite{ginsim}. This package debuted in the mid-2000s from academic descendants of Ren\'e Thomas, and it still remains popular today. However, if a faculty member in charge of a less established software package moves, funding dries up, or group members get interested in another research topic, these non-commercial software tools can disappear at worst, remain immortalized online at an old GitHub repository, or in the best case scenario, get added as libraries to open-source platforms such as Sage (general), R (statistical), and Macaulay2 or Singular (computational algebraic). Algorithms for neural ideals have recently been written for Sage \cite{petersen2018neural} and Matlab \cite{youngs2015neural}. We will refrain from providing a list of citations to software packages on broad topics such as Boolean networks or phylogenetics, and instead encourage the interested reader to do a simple Google search, as the results are enormous.  

In addition to the question about textbooks, another challenge with teaching certain topics in algebraic biology, especially those involving Boolean models, is the abundance of frameworks and the lack of a standard notation. For example, a research group in an electrical engineering department might view a gene regulatory network as a series of logical gates defined by truth tables, and write their code accordingly, whereas others might represent their functions as polynomials over $\F_2$. Some modelers might update their functions synchronously, whereas others use some sort of asynchronous update. Many others use a hybrid scheme, block-sequential update, or introduce some amount of stochasticity or varying time delays. Of course, the other side of the coin to this is that it means that these methods have grown in popularity and are being used by researchers from a diverse range of fields, such as math, computer science, engineering, biology, and physics. 

\section{Concluding remarks}

In order to highlight, promote, and legitimize the relatively unknown field of algebraic biology, we have given a peek at four biological problems whose analysis is amenable to algebraic methods. Most of the thrust of this field has come since the turn of the century, and interest shows no signs of dying down. One of the main questions we addressed in this article is what role it can and should have in the classroom. It is unrealistic to expect that many institutions will offer an entire class on this topic, and one can make a strong case that there are much more appropriate topics for a mathematical biology or modeling class. However, topics from algebraic biology wrapped up in self-contained modules can inspire students not only in traditional modeling or mathematical biology courses, but also in modern algebra courses. Just as questions from physics and biology play a motivating role in calculus courses, biological questions can join the ranks of applications shown to students in their first abstract algebra course. It is not uncommon for mathematicians to motivate a discussion of group theory using questions of symmetry, or to showcase cryptography or coding theory as an application of finite fields. One could also introduce Boolean models or place fields in neuroscience as motivators for understanding polynomial ideals. Computational algebraic techniques such as Gr\"obner bases can be motivated by solving systems of equations from a biological problem -- perhaps a Boolean model of a molecular network or a nonlinear system of differential equations from a biochemical reaction network. 

We will conclude by emphasizing once again that this article is certainly an incomplete and biased glimpse into the world of algebraic biology. There are many topics, individuals, and papers that were not included. Some of this was simply for space limitations -- the number of citations easily could have been an order of magnitude greater, but at some point we had to draw a sensible line. Beyond that, we are just two individuals who have our own natural implicit biases and blind spots, so there are certainly unintentional omissions as well. We encourage readers to reach out and inform us of any such oversights, tangentially related work, or new developments going forward. If enough feedback is given, we may update this article on the arXiv as we see fit. We would love to hear anecdotes from readers about opportunities, experiences, and lessons learned from introducing algebraic biology into their classrooms.

\section*{Acknowledgements}

The authors would like to thank Elena Dimitrova, Heather Harrington, Reinhard Laubenbacher, and Raina Robeva for  their feedback on an earlier draft of this article.

\bibliographystyle{abbrv}
\bibliography{mathbio}

\end{document}